\documentclass[%
 reprint,
showpacs,preprintnumbers,
 amsmath,amssymb,
prb,
floatfix,
]{revtex4-1}

\usepackage{graphicx}
\usepackage{dcolumn}
\usepackage{bm}

\begin{document}

\newcommand{\SVFA}{Sr$_2$VO$_3$FeAs}


\title{Weak magnetism and the Mott-state of vanadium in superconducting {\SVFA}}

\author{Franziska Hummel$^1$}
\author{Yixi Su$^2$}
\author{Anatoliy Senyshyn$^3$}
\author{Dirk Johrendt$^1$}
\email{johrendt@lmu.de}
\affiliation{
{$^1$}{Department Chemie, Ludwig-Maximilians-Universit\"{a}t M\"{u}nchen,\\ Butenandtstr. 5-13 (Haus D), 81377 M\"{u}nchen, Germany}\\
{$^2$}{J\"{u}lich Centre for Neutron Science JCNS-FRM II, Forschungszentrum J\"{u}lich GmbH, Outstation at FRM II, Lichtenbergstrasse 1, 85747 Garching, Germany} \\
{$^3$}{Forschungsneutronenquelle Heinz-Maier-Leibnitz (FRM II), Technische Universit\"{a}t M\"{u}nchen, Lichtenbergstrasse 1, 85747 Garching, Germany}\\
 }

\date{\today}

\begin{abstract}

We report neutron scattering data and DFT calculations of the stoichiometric iron-arsenide superconductor {\SVFA}. Rietveld refinements of neutron powder patterns confirm the ideal composition without oxygen deficiencies. Experiments with polarized neutrons prove weak magnetic ordering in the V-sublattice of {\SVFA} at $\approx$~45~K with a probable propagation vector \textbf{q}~=~($\frac{1}{8},\frac{1}{8}$,0). The ordered moment of $\approx$ 0.1~$\mu_{\rm B}$ is too small to remove the V-3$d$ bands from the Fermi level by magnetic exchange splitting, and much smaller than predicted from a recent LDA+U study. By using DFT calculations with a GGA+EECE functional we recover the typical quasi-nested Fermi-surface even without magnetic moment. From this we suggest that the V-atoms are in a Mott-state where the electronic correlations are dominated by on-site Coulomb-repulsion which shifts the V-3$d$ states away from the Fermi energy. Our results are consistent with photoemission data and clearly reveal that {\SVFA} is a typical iron-arsenide superconductor with quasi-nested hole- and electron-like Fermi surface sheets, and constitutes no new paradigm. We suggest that intrinsic electron-doping through V$^{3+}$/V$^{4+}$-mixed valence is responsible for the absence of SDW ordering.

\end{abstract}



\maketitle


\section{Introduction}

Superconductivity in iron-arsenide compounds occurs in the proximity of magnetic ordering, \cite{Stewart-2011,Johrendt-2011} and it is today believed that magnetic fluctuations play an essential role in the formation of the cooper pairs .\cite{Mazin-2009,Dai-2012,Scalapino-2012} In typical parent compounds like LaFeAsO or BaFe$_2$As$_2$,\cite{Rotter-2008} the stripe-type antiferromagnetic (AF) order of the iron moments becomes destabilized either by doping or applying pressure while superconductivity emerges.\cite{Kamihara-2008,Rotter-2009,Alireza-2009} High critical temperatures above 30~K are obviously confined to this scenario, since other iron-pnictide superconductors like LaFePO \cite{Kamihara-2006} or KFe$_2$As$_2$ \cite{Sasmal-2008} without magnetic ordering exhibit critical temperatures below 5~K. An exception seems to be the iron-arsenide {\SVFA} which is superconducting up to 37~K \cite{Zhu-2009} and even 45~K under pressure \cite{Kotegawa-2009} in its stoichiometric parent phase. No magnetic ordering in the FeAs-layer or structural phase transitions have been found up to now. However, the true stoichiometry of this compound has been a subject of discussion. It was also debated whether intrinsic electron doping through oxygen deficiency or V$^{3+}$/V$^{4+}$ mixed valence might play a role. \cite{Han-2010, Cao-2010} The latter may explain the absence of magnetic ordering, however, the concrete oxygen content has never been determined experimentally, and also the presence of V$^{4+}$ is still unproved.

Shortly after the discovery of {\SVFA} it was suspected that this material may constitute a new paradigm, as its electronic structure did not seem to fulfill the Fermi surface (FS) quasi-nesting condition, which is believed to be essential for high critical temperatures. Indeed first GGA-calculations predicted half-metallic V-3$d$ bands and magnetic properties inconsistent with experimental data.\cite{Shein-2009} Weak hybridization between the V-3$d$ and Fe-3$d$ bands near $E_{\rm{F}}$ changes the FS topology significantly and destroys the quasi-nesting. \cite{Shein-2009,Pickett-2010} However, by weighting the electronic states of this more complex FS with their Fe character, a FS was obtained that again supports the quasi-nesting model.\cite{Mazin-2010} The latter was also confirmed by angle resolved photoemission experiments with {\SVFA} crystals.\cite{Qian-2011} It was suggested that magnetic ordering of the V-sublattice could remove the V-3$d$-states from the Fermi level by magnetic exchange splitting, so as to maintain the characteristic quasi-nested FS topology. First evidence for such magnetic ordering was found by neutron-scattering,\cite{Tegel-2010} but the magnetic structure remained unclear. Antiferromagnetic ordering of the V-sublattice was also indicated by $\mu$SR- and NMR-spectroscopy.\cite{Munevar-2011} LDA+U calculations revealed an antiferromagnetic ground state with a gap in the V-3$d$ bands and a magnetic moment of $\approx$1.8~$\mu_{\rm B}$ per vanadium. \cite{Nakamura-2010} In this case, the magnetic exchange splitting would be large enough to open a significant gap in the V-3$d$ bands, but on the other hand, a magnetic moment of such magnitude has not been observed by any experiment so far.

In this article we present neutron diffraction data of polycrystalline {\SVFA} and DFT calculations to understand the results. By using polarized neutrons we  show that antiferromagnetic ordering evolves around 45~K with a magnetic moment of $\approx$ 0.1~$\mu_{\rm B}$ per vanadium, thus more than one order of magnitude smaller than the LDA+U prediction. GGA+EECE calculations reveal that the V-3$d$ orbitals are in a Mott-state and subject to strong on-site repulsion correlations which remove the V-states from the Fermi level without significant magnetic moment, leading to a consistent picture of this superconducting material.

\section{Experimental details}

A polycrystalline sample of {\SVFA} (4~g) was synthesized in four separate 1~g batches. Stoichiometric mixtures of Sr, V, Fe$_2$O$_3$ and As$_2$O$_3$ were transferred into alumina crucibles and sealed in quartz ampoules under argon atmosphere. The samples were heated up to 1323~K for 60~h at rates of 60~K/h and 200~K/h for heating and cooling, respectively. Afterwards, the samples were ground in agate mortars, pressed into pellets and sintered for 60~h at 1323~K. The batches were then united, homogenized, pressed into two pellets, both pellets being placed in the same alumina crucible, and sintered again for 60~h at 1323~K. The product was obtained as a black, polycrystalline and air-stable sample.
Room temperature X-Ray powder diffraction patterns were recorded using a STOE Stadi P (Cu-$K_{\alpha 1}$ radiation). Room temperature high resolution neutron diffraction data were measured at SPODI (FRM II, Garching, Germany) with an incident wavelength of 0.1548~nm. Measurements with polarized neutrons were recorded at the polarized neutron spectrometer DNS with an incident wavelength of 0.42~nm. Polarization analysis was performed via the xyz-method.\cite{xyz} For Rietveld refinements of the data the TOPAS package \cite{Topas-2007} was used with the fundamental parameter approach as reflection profiles. March Dollase or spherical harmonics functions were used to describe the preferred orientation of the crystallites. Magnetic measurements were performed using a SQUID magnetometer (MPMS-XL5, Quantum Design, Inc.).

Electronic band structure calculations were performed with the WIEN2k program package \cite{Blaha-2001} using density functional theory (DFT) within the full-potential LAPW-lo method and the generalized gradient approximation (GGA) with a separation energy for core and valence states of 6~Ry. The energy and charge convergence criteria were chosen to be 10$^{-5}$~Ry/cell and 10$^{-4}$~ e/cell, respectively, and 56 irreducible k-points were used with a cutoff for plane waves R$_{mt}$K$_{max}$~=~7.0. For detailed descriptions see \cite{Schwarz-2003} and \cite{Singh-2006}. In order to reproduce the Mott insulating state of the vanadium-oxide layer in {\SVFA} we have used the EECE (\underline{E}xact \underline{E}xchange of \underline{C}orrelated \underline{E}lectrons) approach implemented in the package.\cite{Novak-2006} The functional is obtained by removing the local exchange-correlation energy of the V-3$d$ orbitals and replacing it with the exact Hartree-Fock (HF) exchange energy. Thus the exchange term is corrected by an exact expression instead of approximations as in LDA+U or GGA+U schemes.

\section{Results and discussion}

\begin{figure}
\center{
\includegraphics[width=90mm]{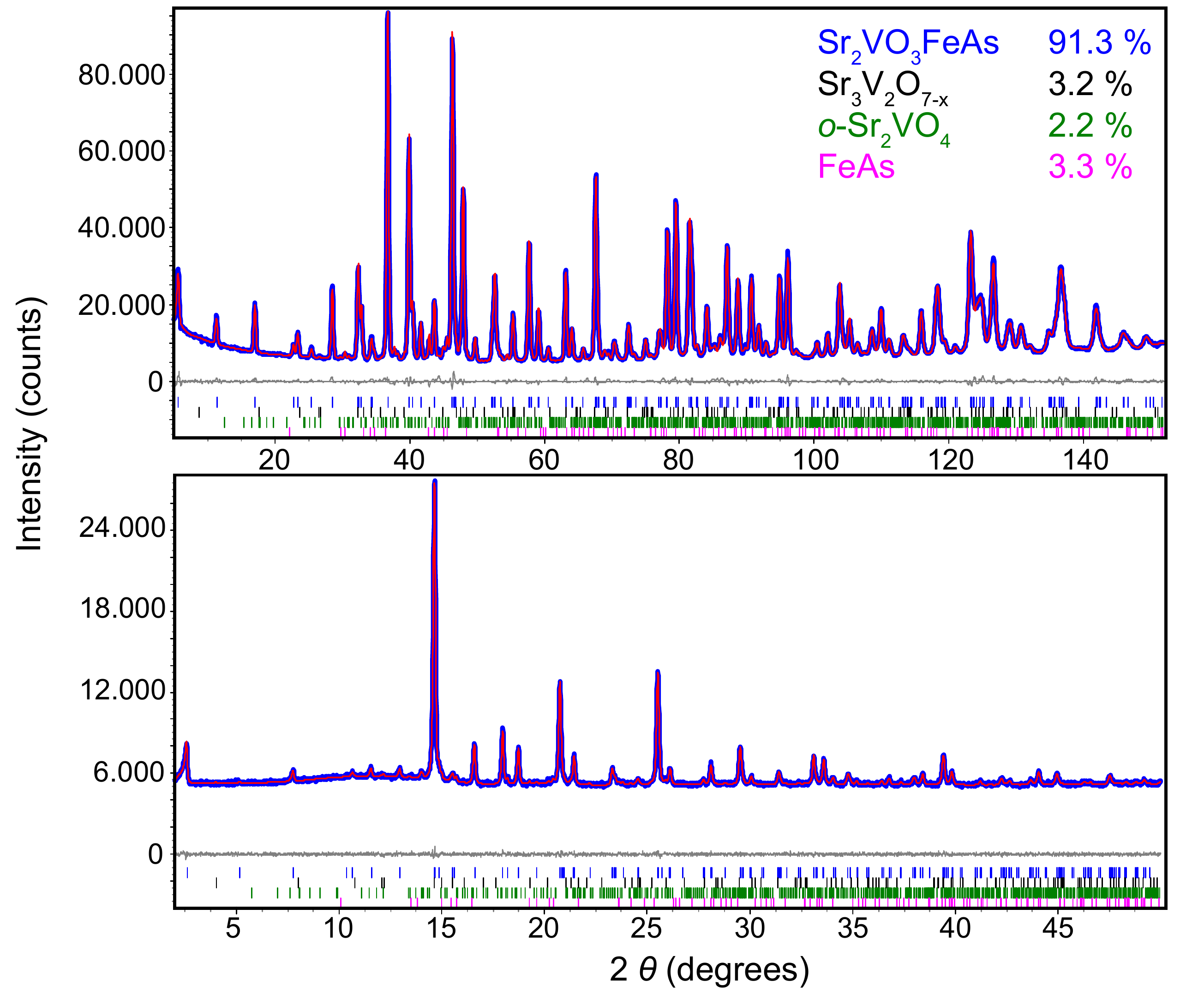}
\caption{X-Ray (bottom) and SPODI (top) neutron diffraction pattern of {\SVFA} with Rietveld refinements.}
\label{fig:spodi}
}
\end{figure}

Rietveld refinements of the X-ray and SPODI neutron data (Fig.~\ref{fig:spodi}) revealed the sample composition 91.3~\% {\SVFA} with 3.2~\% Sr$_3$V$_2$O$_{7-x}$, 2.2~\% orthorhombic Sr$_2$VO$_4$, and 3.3~\% FeAs as impurity phases ($wt$-\%). No oxygen deficiency was detected in the {\SVFA}-phase, since both oxygen sites remained fully occupied within the experimental errors during the refinement. Also, no vanadium was detected on the iron site, hence, the {\SVFA} phase in the sample turns out to be stoichiometric.

\begin{figure}
\center{
\includegraphics[width=80mm]{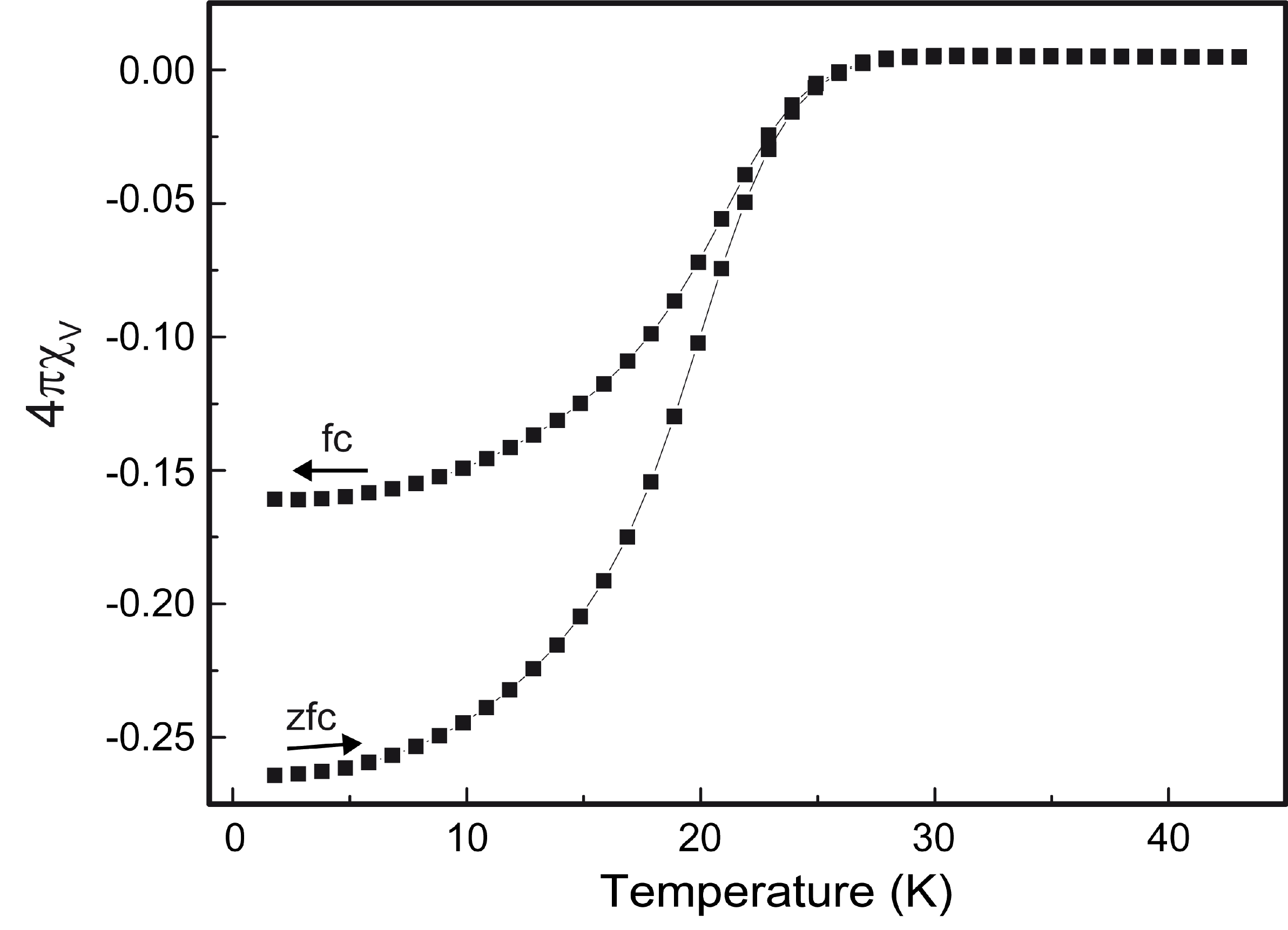}
\caption{Zero-field-cooled/field-cooled susceptibility of {\SVFA} at 15 Oe.}
\label{fig:chi}
}
\end{figure}

Magnetic susceptibility measurements (Fig.~\ref{fig:chi}) revealed that the sample is superconducting with $T_c$ = 25~K. Thus, our stoichiometric {\SVFA} has a comparably low $T_c$, which contradicts to the work of Han \textit{et al.} where nominally stoichiometric samples exhibit the highest critical temperatures and decreasing values were attributed to oxygen deficiencies.\cite{Han-2010} The superconducting volume fraction of our sample is roughly estimated to be about 26~\%. Although most experimental works do not mention the superconducting volume fractions achieved, such small values have been reported before \cite{Tegel-2010} and seem to be another general peculiarity in this system besides the strongly differing superconducting transition temperatures.

\begin{figure}
\center{
\includegraphics[width=80mm]{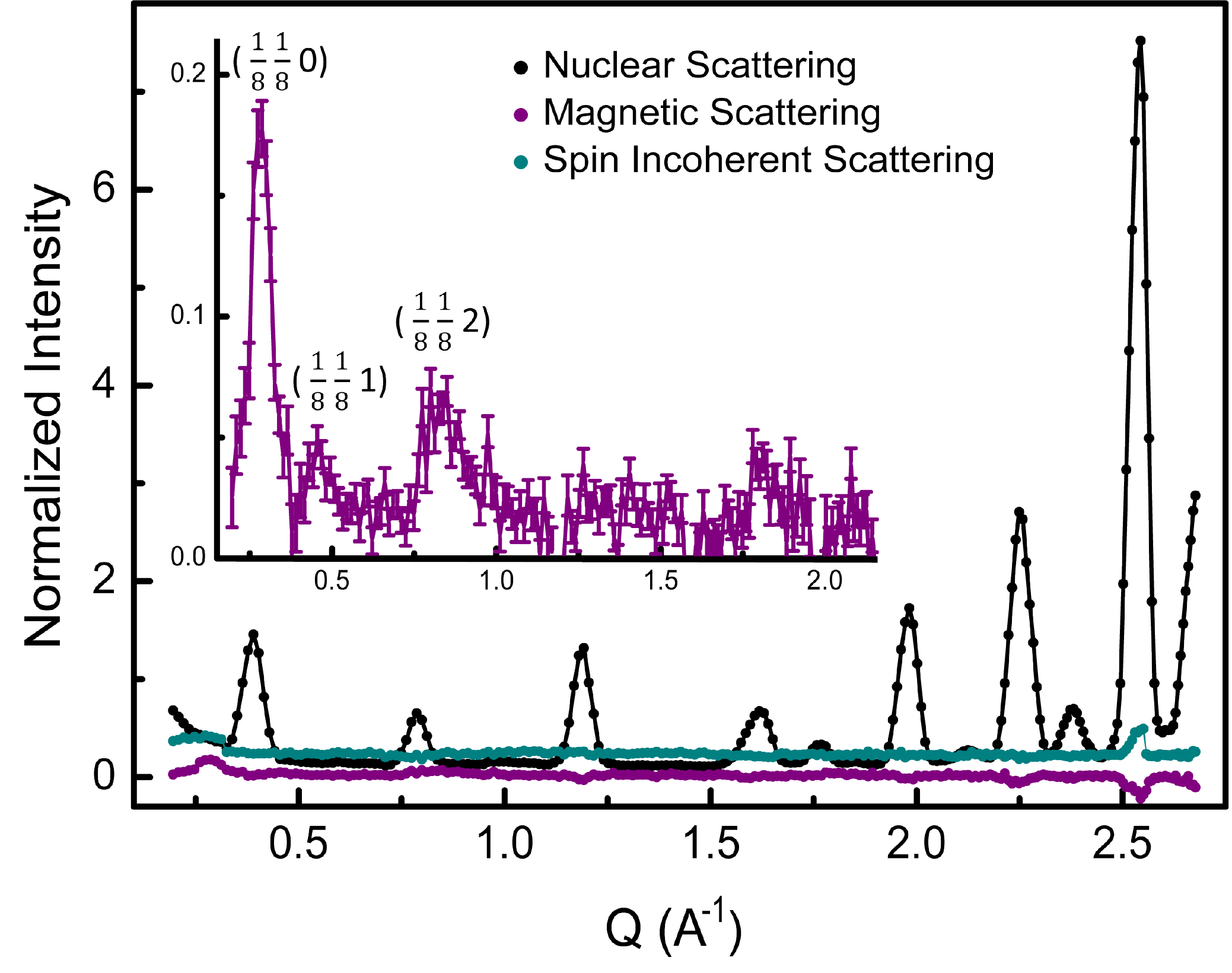}
\caption{DNS neutron scattering pattern of {\SVFA} at 2~K divided into nuclear (black), magnetic (purple) and spin incoherent (green) contributions. Inset: Enlarged image of the magnetic contribution.}
\label{fig:dns}
}
\end{figure}

Our earlier neutron diffraction experiments revealed weak additional reflections at low temperatures which indicate possible magnetic ordering of vanadium.\cite{Tegel-2010} In order to clarify the origin of these reflections, we performed low temperature diffraction experiments with polarized neutrons. Hereby, nuclear and magnetic reflections can be separated from each other and from the spin incoherent scattering. Figure~\ref{fig:dns} shows the intensity relation between all three contributions. All intensities have been normalized to the total spin incoherent scattering cross-section of vanadium. It is evident that the magnetic scattering is very small compared to the nuclear part. Nevertheless, significant magnetic peaks are visible (inset of Fig.~\ref{fig:dns}).

\begin{figure}
\center{
\includegraphics[width=80mm]{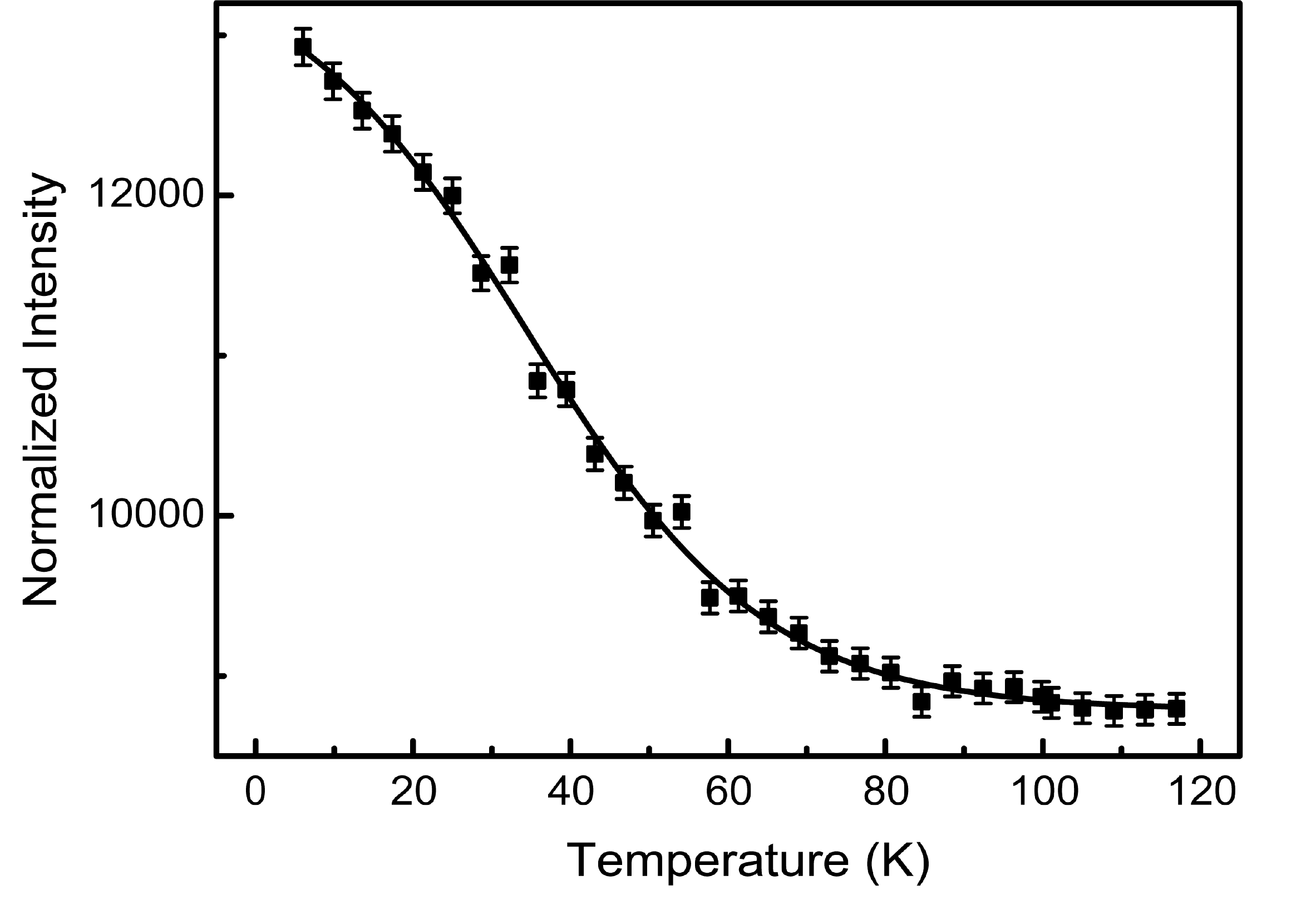}
\caption{Temperature dependency of the magnetic ordering parameter of the V-sublattice in {\SVFA} measured at $Q$~=~0.29~{\AA}$^{-1}$.}
\label{fig:temp}
}
\end{figure}

These reflections differ from those observed earlier \cite{Tegel-2010} because of the spin incoherent scattering contribution. Due to the large incoherent scattering length of $^{51}$V this contribution causes a strong scattering background.  The polarization analysis allows to filter out the true magnetic reflections shown in the inset of Figure~\ref{fig:dns}. These reflections can be indexed as ($\frac{1}{8},\frac{1}{8}$,l) and thus point to a magnetic ordering of the vanadium moments with a propagation vector \textbf{q} = ($\frac{1}{8},\frac{1}{8}$,0). Due to different intensities of the reflections, we assume that the magnetic moments are aligned along [001].
The temperature dependency of the magnetic ordering parameter is shown in Figure~\ref{fig:temp}. The magnetic transition at about 45~K is consistent with magnetic susceptibility measurements \cite{Tegel-2010,Cao-2010} which showed anomalies around 50~K which is related to the magnetic transition reported here. However, our neutron experiments showed no significant change of the diffraction pattern in the temperature range of the further reported susceptibility anomalies around 70~K and 150~K.

\begin{figure}
\center{
\includegraphics[width=90mm]{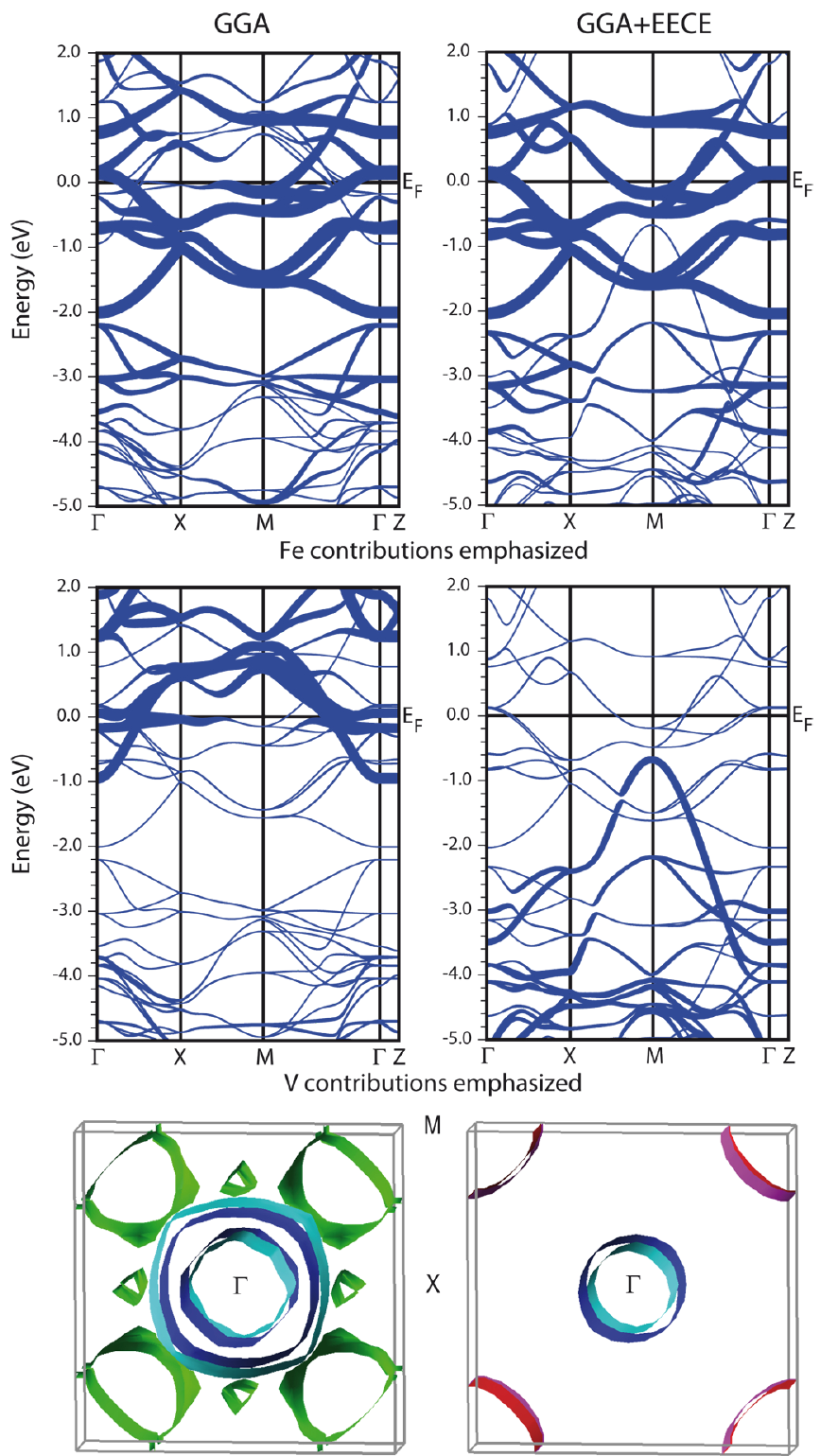}
\caption{Band structure of {\SVFA} with iron (top) and vanadium (middle) contributions emphasized and Fermi surface (bottom) calculated with GGA (left) and GGA+EECE (right).}
\label{fig:dft}
}
\end{figure}

From the magnetic scattering intensity we estimate an ordered magnetic moment of $\approx$ 0.1~$\mu_{\rm B}$/V in agreement with the $\mu$SR experiments,\cite{Munevar-2011} but in stark contrast to LDA+U calculations which predicted  much higher values around 1.8~$\mu_{\rm B}$/V.\cite{Nakamura-2010} However, the magnetic exchange splitting of a 0.1~$\mu_{\rm B}$ magnetic moment is certainly too small to remove the V-$3d$ states from the Fermi energy. We therefore suggest that the V-atoms in {\SVFA} are in a Mott-state where the electronic correlation is dominated by the on-site coulomb-repulsion, but without magnetic ordering. In order to check this idea, we performed GGA+EECE calculations with an exact (Hartree-Fock) exchange correction applied to the V-3$d$ orbitals without magnetic moment. Figure~\ref{fig:dft} shows the band structures and Fermi surfaces of non-magnetic {\SVFA} calculated with standard GGA and with GGA+EECE, respectively. The standard GGA calculation produces metallic V-3$d$ bands, and consequently a FS which is more complex due to the hybridization of V-3$d$ with Fe-3$d$ bands. By applying the EECE functional the V-3$d$ bands are shifted to much lower energies, and the vicinity of the Fermi energy is now dominated by the Fe-3$d$ states as known from other iron-arsenide materials. Even without magnetic moment, the FS shows the quasi-nesting between hole-like and electron-like cylinders, which is the typical FS-topology of iron-arsenide superconductors. LDA+DMFT calculations provide similar results.\cite{Aichhorn-2013} However, the absence of magnetic ordering in the Fe-sublattice and superconductivity in the stoichiometric phase indicates that {\SVFA} is intrinsically doped, probably by a certain degree of V$^{3+}$/V$^{4+}$ mixed valence.

\section{Conclusion}

Our results consistently show that {\SVFA} is a typical iron-arsenide superconductor with quasi-nested FS topology, and constitutes no new paradigm. No deviations from the ideal composition were found by Rietveld-refinements of neutron powder patterns, ruling out oxygen deficiency as well as V-doping at the Fe site. Magnetic ordering in the V-sublattice was detected by polarized neutron scattering with a probable propagation vector \textbf{q} = ($\frac{1}{8},\frac{1}{8}$,0). The estimated ordered moment of 0.1~$\mu_{\rm B}$/V is much smaller than recently predicted by LDA-U calculations, and certainly too small to produce a significant gap in the V-3$d$ bands by magnetic exchange splitting. However, our DFT calculations using a GGA+EECE functional revealed the typical quasi-nested FS even without magnetic moment. The V-atoms are in a Mott-state where the electronic correlations are dominated by on-site Coulomb-repulsion, which shifts the V-3$d$ states away from the Fermi energy. We suggest that intrinsic electron-doping through V$^{3+}$/V$^{4+}$-mixed valence is responsible for the absence of SDW ordering and thus for superconductivity even in the undoped phase.

\begin{acknowledgments}

The authors would like to thank Dr. Markus Aichhorn for his advice on the theoretical calculations. This work was financially supported by the German Research Foundation (DFG) within the priority program SPP1458, Project No. JO257/6-2.

\end{acknowledgments}

\bibliographystyle{apsrev}

\bibliography{V21311_PRB}

\begin{thebibliography}{29}
\expandafter\ifx\csname natexlab\endcsname\relax\def\natexlab#1{#1}\fi
\expandafter\ifx\csname bibnamefont\endcsname\relax
  \def\bibnamefont#1{#1}\fi
\expandafter\ifx\csname bibfnamefont\endcsname\relax
  \def\bibfnamefont#1{#1}\fi
\expandafter\ifx\csname citenamefont\endcsname\relax
  \def\citenamefont#1{#1}\fi
\expandafter\ifx\csname url\endcsname\relax
  \def\url#1{\texttt{#1}}\fi
\expandafter\ifx\csname urlprefix\endcsname\relax\def\urlprefix{URL }\fi
\providecommand{\bibinfo}[2]{#2}
\providecommand{\eprint}[2][]{\url{#2}}

\bibitem[{\citenamefont{Stewart}(2011)}]{Stewart-2011}
\bibinfo{author}{\bibfnamefont{G.~R.} \bibnamefont{Stewart}},
  \bibinfo{journal}{Rev. Mod. Phys.} \textbf{\bibinfo{volume}{83}},
  \bibinfo{pages}{1589} (\bibinfo{year}{2011}).

\bibitem[{\citenamefont{Johrendt}(2011)}]{Johrendt-2011}
\bibinfo{author}{\bibfnamefont{D.}~\bibnamefont{Johrendt}},
  \bibinfo{journal}{J. Mater. Chem.} \textbf{\bibinfo{volume}{21}},
  \bibinfo{pages}{13726} (\bibinfo{year}{2011}).

\bibitem[{\citenamefont{Mazin and Johannes}(2009)}]{Mazin-2009}
\bibinfo{author}{\bibfnamefont{I.~I.} \bibnamefont{Mazin}} \bibnamefont{and}
  \bibinfo{author}{\bibfnamefont{M.~D.} \bibnamefont{Johannes}},
  \bibinfo{journal}{Nature Physics} \textbf{\bibinfo{volume}{5}},
  \bibinfo{pages}{141} (\bibinfo{year}{2009}).

\bibitem[{\citenamefont{Dai et~al.}(2012)\citenamefont{Dai, Hu, and
  Dagotto}}]{Dai-2012}
\bibinfo{author}{\bibfnamefont{P.}~\bibnamefont{Dai}},
  \bibinfo{author}{\bibfnamefont{J.}~\bibnamefont{Hu}}, \bibnamefont{and}
  \bibinfo{author}{\bibfnamefont{E.}~\bibnamefont{Dagotto}},
  \bibinfo{journal}{Nat Phys} \textbf{\bibinfo{volume}{8}},
  \bibinfo{pages}{709} (\bibinfo{year}{2012}).

\bibitem[{\citenamefont{Scalapino}(2012)}]{Scalapino-2012}
\bibinfo{author}{\bibfnamefont{D.~J.} \bibnamefont{Scalapino}},
  \bibinfo{journal}{Rev. Mod. Phys.} \textbf{\bibinfo{volume}{84}},
  \bibinfo{pages}{1383} (\bibinfo{year}{2012}).

\bibitem[{\citenamefont{Rotter et~al.}(2008{\natexlab{a}})\citenamefont{Rotter,
  Tegel, Schellenberg, Hermes, P\"{o}ttgen, and Johrendt}}]{Rotter-2008}
\bibinfo{author}{\bibfnamefont{M.}~\bibnamefont{Rotter}},
  \bibinfo{author}{\bibfnamefont{M.}~\bibnamefont{Tegel}},
  \bibinfo{author}{\bibfnamefont{I.}~\bibnamefont{Schellenberg}},
  \bibinfo{author}{\bibfnamefont{W.}~\bibnamefont{Hermes}},
  \bibinfo{author}{\bibfnamefont{R.}~\bibnamefont{P\"{o}ttgen}},
  \bibnamefont{and} \bibinfo{author}{\bibfnamefont{D.}~\bibnamefont{Johrendt}},
  \bibinfo{journal}{Phys. Rev. B} \textbf{\bibinfo{volume}{78}},
  \bibinfo{pages}{020503(R)} (\bibinfo{year}{2008}{\natexlab{a}}).

\bibitem[{\citenamefont{Kamihara et~al.}(2008)\citenamefont{Kamihara, Watanabe,
  Hirano, and Hosono}}]{Kamihara-2008}
\bibinfo{author}{\bibfnamefont{Y.}~\bibnamefont{Kamihara}},
  \bibinfo{author}{\bibfnamefont{T.}~\bibnamefont{Watanabe}},
  \bibinfo{author}{\bibfnamefont{M.}~\bibnamefont{Hirano}}, \bibnamefont{and}
  \bibinfo{author}{\bibfnamefont{H.}~\bibnamefont{Hosono}},
  \bibinfo{journal}{J. Am. Chem. Soc.} \textbf{\bibinfo{volume}{130}},
  \bibinfo{pages}{3296} (\bibinfo{year}{2008}).

\bibitem[{\citenamefont{Rotter et~al.}(2008{\natexlab{b}})\citenamefont{Rotter,
  Tegel, and Johrendt}}]{Rotter-2009}
\bibinfo{author}{\bibfnamefont{M.}~\bibnamefont{Rotter}},
  \bibinfo{author}{\bibfnamefont{M.}~\bibnamefont{Tegel}}, \bibnamefont{and}
  \bibinfo{author}{\bibfnamefont{D.}~\bibnamefont{Johrendt}},
  \bibinfo{journal}{Phys. Rev. Lett.} \textbf{\bibinfo{volume}{101}},
  \bibinfo{pages}{107006} (\bibinfo{year}{2008}{\natexlab{b}}).

\bibitem[{\citenamefont{Alireza et~al.}(2009)\citenamefont{Alireza, Ko,
  Gillett, Petrone, Cole, Lonzarich, and Sebastian}}]{Alireza-2009}
\bibinfo{author}{\bibfnamefont{P.~L.} \bibnamefont{Alireza}},
  \bibinfo{author}{\bibfnamefont{Y.~T.~C.} \bibnamefont{Ko}},
  \bibinfo{author}{\bibfnamefont{J.}~\bibnamefont{Gillett}},
  \bibinfo{author}{\bibfnamefont{C.~M.} \bibnamefont{Petrone}},
  \bibinfo{author}{\bibfnamefont{J.~M.} \bibnamefont{Cole}},
  \bibinfo{author}{\bibfnamefont{G.~G.} \bibnamefont{Lonzarich}},
  \bibnamefont{and} \bibinfo{author}{\bibfnamefont{S.~E.}
  \bibnamefont{Sebastian}}, \bibinfo{journal}{J. Phys.: Condens. Matter}
  \textbf{\bibinfo{volume}{21}}, \bibinfo{pages}{012208}
  (\bibinfo{year}{2009}).

\bibitem[{\citenamefont{Kamihara et~al.}(2006)\citenamefont{Kamihara,
  Hiramatsu, Hirano, Kawamura, Yanagi, Kamiya, and Hosono}}]{Kamihara-2006}
\bibinfo{author}{\bibfnamefont{Y.}~\bibnamefont{Kamihara}},
  \bibinfo{author}{\bibfnamefont{H.}~\bibnamefont{Hiramatsu}},
  \bibinfo{author}{\bibfnamefont{M.}~\bibnamefont{Hirano}},
  \bibinfo{author}{\bibfnamefont{R.}~\bibnamefont{Kawamura}},
  \bibinfo{author}{\bibfnamefont{H.}~\bibnamefont{Yanagi}},
  \bibinfo{author}{\bibfnamefont{T.}~\bibnamefont{Kamiya}}, \bibnamefont{and}
  \bibinfo{author}{\bibfnamefont{H.}~\bibnamefont{Hosono}},
  \bibinfo{journal}{J. Am. Chem. Soc.} \textbf{\bibinfo{volume}{128}},
  \bibinfo{pages}{10012} (\bibinfo{year}{2006}).

\bibitem[{\citenamefont{Sasmal et~al.}(2008)\citenamefont{Sasmal, Lv, Lorenz,
  Guloy, Chen, Xue, and Chu}}]{Sasmal-2008}
\bibinfo{author}{\bibfnamefont{K.}~\bibnamefont{Sasmal}},
  \bibinfo{author}{\bibfnamefont{B.}~\bibnamefont{Lv}},
  \bibinfo{author}{\bibfnamefont{B.}~\bibnamefont{Lorenz}},
  \bibinfo{author}{\bibfnamefont{A.}~\bibnamefont{Guloy}},
  \bibinfo{author}{\bibfnamefont{F.}~\bibnamefont{Chen}},
  \bibinfo{author}{\bibfnamefont{Y.}~\bibnamefont{Xue}}, \bibnamefont{and}
  \bibinfo{author}{\bibfnamefont{C.~W.} \bibnamefont{Chu}},
  \bibinfo{journal}{Phys. Rev. Lett.} \textbf{\bibinfo{volume}{101}},
  \bibinfo{pages}{107007} (\bibinfo{year}{2008}).

\bibitem[{\citenamefont{Zhu et~al.}(2009)\citenamefont{Zhu, Han, Mu, Cheng,
  Shen, Zeng, and Wen}}]{Zhu-2009}
\bibinfo{author}{\bibfnamefont{X.}~\bibnamefont{Zhu}},
  \bibinfo{author}{\bibfnamefont{F.}~\bibnamefont{Han}},
  \bibinfo{author}{\bibfnamefont{G.}~\bibnamefont{Mu}},
  \bibinfo{author}{\bibfnamefont{P.}~\bibnamefont{Cheng}},
  \bibinfo{author}{\bibfnamefont{B.}~\bibnamefont{Shen}},
  \bibinfo{author}{\bibfnamefont{B.}~\bibnamefont{Zeng}}, \bibnamefont{and}
  \bibinfo{author}{\bibfnamefont{H.-H.} \bibnamefont{Wen}},
  \bibinfo{journal}{Phys. Rev. B} \textbf{\bibinfo{volume}{79}},
  \bibinfo{pages}{220512} (\bibinfo{year}{2009}).

\bibitem[{\citenamefont{Kotegawa et~al.}(2009)\citenamefont{Kotegawa, Kawazoe,
  Tou, Murata, Ogino, Kishio, and Shimoyama}}]{Kotegawa-2009}
\bibinfo{author}{\bibfnamefont{H.}~\bibnamefont{Kotegawa}},
  \bibinfo{author}{\bibfnamefont{T.}~\bibnamefont{Kawazoe}},
  \bibinfo{author}{\bibfnamefont{H.}~\bibnamefont{Tou}},
  \bibinfo{author}{\bibfnamefont{K.}~\bibnamefont{Murata}},
  \bibinfo{author}{\bibfnamefont{H.}~\bibnamefont{Ogino}},
  \bibinfo{author}{\bibfnamefont{K.}~\bibnamefont{Kishio}}, \bibnamefont{and}
  \bibinfo{author}{\bibfnamefont{J.}~\bibnamefont{Shimoyama}},
  \bibinfo{journal}{J. Phys. Soc. Jpn.} \textbf{\bibinfo{volume}{78}},
  \bibinfo{pages}{123707} (\bibinfo{year}{2009}).

\bibitem[{\citenamefont{Han et~al.}(2010)\citenamefont{Han, Zhu, Mu, Cheng,
  Shen, Zeng, and Wan}}]{Han-2010}
\bibinfo{author}{\bibfnamefont{F.}~\bibnamefont{Han}},
  \bibinfo{author}{\bibfnamefont{X.}~\bibnamefont{Zhu}},
  \bibinfo{author}{\bibfnamefont{G.}~\bibnamefont{Mu}},
  \bibinfo{author}{\bibfnamefont{P.}~\bibnamefont{Cheng}},
  \bibinfo{author}{\bibfnamefont{B.}~\bibnamefont{Shen}},
  \bibinfo{author}{\bibfnamefont{B.}~\bibnamefont{Zeng}}, \bibnamefont{and}
  \bibinfo{author}{\bibfnamefont{H.}~\bibnamefont{Wan}}, \bibinfo{journal}{Sci.
  China Phys. Mech. Astron.} \textbf{\bibinfo{volume}{53}},
  \bibinfo{pages}{1202} (\bibinfo{year}{2010}).

\bibitem[{\citenamefont{Cao et~al.}(2010)\citenamefont{Cao, Ma, Wang, Sun, Bao,
  Jiang, Luo, Feng, Zhou, Xie et~al.}}]{Cao-2010}
\bibinfo{author}{\bibfnamefont{G.-H.} \bibnamefont{Cao}},
  \bibinfo{author}{\bibfnamefont{Z.}~\bibnamefont{Ma}},
  \bibinfo{author}{\bibfnamefont{C.}~\bibnamefont{Wang}},
  \bibinfo{author}{\bibfnamefont{Y.}~\bibnamefont{Sun}},
  \bibinfo{author}{\bibfnamefont{J.}~\bibnamefont{Bao}},
  \bibinfo{author}{\bibfnamefont{S.}~\bibnamefont{Jiang}},
  \bibinfo{author}{\bibfnamefont{Y.}~\bibnamefont{Luo}},
  \bibinfo{author}{\bibfnamefont{C.}~\bibnamefont{Feng}},
  \bibinfo{author}{\bibfnamefont{Y.}~\bibnamefont{Zhou}},
  \bibinfo{author}{\bibfnamefont{Z.}~\bibnamefont{Xie}}, \bibnamefont{et~al.},
  \bibinfo{journal}{Phys. Rev. B} \textbf{\bibinfo{volume}{82}},
  \bibinfo{pages}{104518} (\bibinfo{year}{2010}).

\bibitem[{\citenamefont{Shein and Ivanovskii}(2009)}]{Shein-2009}
\bibinfo{author}{\bibfnamefont{I.~R.} \bibnamefont{Shein}} \bibnamefont{and}
  \bibinfo{author}{\bibfnamefont{A.~L.} \bibnamefont{Ivanovskii}},
  \bibinfo{journal}{Journal of Superconductivity and Novel Magnetism}
  \textbf{\bibinfo{volume}{22}}, \bibinfo{pages}{613} (\bibinfo{year}{2009}).

\bibitem[{\citenamefont{Lee and Pickett}(2010)}]{Pickett-2010}
\bibinfo{author}{\bibfnamefont{K.~W.} \bibnamefont{Lee}} \bibnamefont{and}
  \bibinfo{author}{\bibfnamefont{W.~E.} \bibnamefont{Pickett}},
  \bibinfo{journal}{Europhys. Lett.} \textbf{\bibinfo{volume}{89}},
  \bibinfo{pages}{57008} (\bibinfo{year}{2010}).

\bibitem[{\citenamefont{Mazin}(2010)}]{Mazin-2010}
\bibinfo{author}{\bibfnamefont{I.~I.} \bibnamefont{Mazin}},
  \bibinfo{journal}{Phys. Rev. B} \textbf{\bibinfo{volume}{81}},
  \bibinfo{pages}{020507} (\bibinfo{year}{2010}).

\bibitem[{\citenamefont{Qian et~al.}(2011)\citenamefont{Qian, Xu, Shi,
  Nakayama, Richard, Kawahara, Sato, Takahashi, Neupane, Xu
  et~al.}}]{Qian-2011}
\bibinfo{author}{\bibfnamefont{T.}~\bibnamefont{Qian}},
  \bibinfo{author}{\bibfnamefont{N.}~\bibnamefont{Xu}},
  \bibinfo{author}{\bibfnamefont{Y.~B.} \bibnamefont{Shi}},
  \bibinfo{author}{\bibfnamefont{K.}~\bibnamefont{Nakayama}},
  \bibinfo{author}{\bibfnamefont{P.}~\bibnamefont{Richard}},
  \bibinfo{author}{\bibfnamefont{T.}~\bibnamefont{Kawahara}},
  \bibinfo{author}{\bibfnamefont{T.}~\bibnamefont{Sato}},
  \bibinfo{author}{\bibfnamefont{T.}~\bibnamefont{Takahashi}},
  \bibinfo{author}{\bibfnamefont{M.}~\bibnamefont{Neupane}},
  \bibinfo{author}{\bibfnamefont{Y.~M.} \bibnamefont{Xu}},
  \bibnamefont{et~al.}, \bibinfo{journal}{Phys. Rev. B}
  \textbf{\bibinfo{volume}{83}}, \bibinfo{pages}{140513}
  (\bibinfo{year}{2011}).

\bibitem[{\citenamefont{Tegel et~al.}(2010)\citenamefont{Tegel, Schmid,
  St\"{u}rzer, Egawa, Su, Senyshyn, and Johrendt}}]{Tegel-2010}
\bibinfo{author}{\bibfnamefont{M.}~\bibnamefont{Tegel}},
  \bibinfo{author}{\bibfnamefont{T.}~\bibnamefont{Schmid}},
  \bibinfo{author}{\bibfnamefont{T.}~\bibnamefont{St\"{u}rzer}},
  \bibinfo{author}{\bibfnamefont{M.}~\bibnamefont{Egawa}},
  \bibinfo{author}{\bibfnamefont{Y.~X.} \bibnamefont{Su}},
  \bibinfo{author}{\bibfnamefont{A.}~\bibnamefont{Senyshyn}}, \bibnamefont{and}
  \bibinfo{author}{\bibfnamefont{D.}~\bibnamefont{Johrendt}},
  \bibinfo{journal}{Phys. Rev. B} \textbf{\bibinfo{volume}{82}},
  \bibinfo{pages}{140507} (\bibinfo{year}{2010}).

\bibitem[{\citenamefont{Munevar et~al.}(2011)\citenamefont{Munevar,
  S\'{a}nchez, Alzamora, Baggio-Saitovitch, Carlo, Goko, Aczel, Williams, Luke,
  Wen et~al.}}]{Munevar-2011}
\bibinfo{author}{\bibfnamefont{J.}~\bibnamefont{Munevar}},
  \bibinfo{author}{\bibfnamefont{D.~R.} \bibnamefont{S\'{a}nchez}},
  \bibinfo{author}{\bibfnamefont{M.}~\bibnamefont{Alzamora}},
  \bibinfo{author}{\bibfnamefont{E.}~\bibnamefont{Baggio-Saitovitch}},
  \bibinfo{author}{\bibfnamefont{J.~P.} \bibnamefont{Carlo}},
  \bibinfo{author}{\bibfnamefont{T.}~\bibnamefont{Goko}},
  \bibinfo{author}{\bibfnamefont{A.~A.} \bibnamefont{Aczel}},
  \bibinfo{author}{\bibfnamefont{T.~J.} \bibnamefont{Williams}},
  \bibinfo{author}{\bibfnamefont{G.~M.} \bibnamefont{Luke}},
  \bibinfo{author}{\bibfnamefont{H.-H.} \bibnamefont{Wen}},
  \bibnamefont{et~al.}, \bibinfo{journal}{Phys. Rev. B}
  \textbf{\bibinfo{volume}{84}}, \bibinfo{pages}{024527}
  (\bibinfo{year}{2011}).

\bibitem[{\citenamefont{Nakamura and Machida}(2010)}]{Nakamura-2010}
\bibinfo{author}{\bibfnamefont{H.}~\bibnamefont{Nakamura}} \bibnamefont{and}
  \bibinfo{author}{\bibfnamefont{M.}~\bibnamefont{Machida}},
  \bibinfo{journal}{Phys. Rev. B} \textbf{\bibinfo{volume}{82}}
  (\bibinfo{year}{2010}).

\bibitem[{\citenamefont{Schärpf and Capellmann}(1993)}]{xyz}
\bibinfo{author}{\bibfnamefont{O.}~\bibnamefont{Schärpf}} \bibnamefont{and}
  \bibinfo{author}{\bibfnamefont{H.}~\bibnamefont{Capellmann}},
  \bibinfo{journal}{Physica Status Solidi (A)} \textbf{\bibinfo{volume}{135}},
  \bibinfo{pages}{359} (\bibinfo{year}{1993}).

\bibitem[{\citenamefont{Coelho}(2007)}]{Topas-2007}
\bibinfo{author}{\bibfnamefont{A.}~\bibnamefont{Coelho}},
  \emph{\bibinfo{title}{TOPAS-Academic, Version 4.1, Coelho Software}}
  (\bibinfo{address}{Brisbane}, \bibinfo{year}{2007}).

\bibitem[{\citenamefont{Blaha et~al.}(2001)\citenamefont{Blaha, Schwarz,
  Madsen, Kvasnicka, and Luitz}}]{Blaha-2001}
\bibinfo{author}{\bibfnamefont{P.}~\bibnamefont{Blaha}},
  \bibinfo{author}{\bibfnamefont{K.}~\bibnamefont{Schwarz}},
  \bibinfo{author}{\bibfnamefont{G.~K.~H.} \bibnamefont{Madsen}},
  \bibinfo{author}{\bibfnamefont{D.}~\bibnamefont{Kvasnicka}},
  \bibnamefont{and} \bibinfo{author}{\bibfnamefont{J.}~\bibnamefont{Luitz}},
  \emph{\bibinfo{title}{Wien2k - an augmented plane wave + local orbitals
  program for calculating crystal properties}} (\bibinfo{year}{2001}).

\bibitem[{\citenamefont{Schwarz and Blaha}(2003)}]{Schwarz-2003}
\bibinfo{author}{\bibfnamefont{K.}~\bibnamefont{Schwarz}} \bibnamefont{and}
  \bibinfo{author}{\bibfnamefont{P.}~\bibnamefont{Blaha}},
  \bibinfo{journal}{Comput. Mat. Sci.} \textbf{\bibinfo{volume}{28}},
  \bibinfo{pages}{259} (\bibinfo{year}{2003}).

\bibitem[{\citenamefont{Singh and Nordstrom}(2006)}]{Singh-2006}
\bibinfo{author}{\bibfnamefont{D.~J.} \bibnamefont{Singh}} \bibnamefont{and}
  \bibinfo{author}{\bibfnamefont{L.}~\bibnamefont{Nordstrom}},
  \emph{\bibinfo{title}{Planewaves, Pseudopotentials and the LAPW Method}}
  (\bibinfo{publisher}{Springer}, \bibinfo{address}{New York},
  \bibinfo{year}{2006}).

\bibitem[{\citenamefont{Novak et~al.}(2006)\citenamefont{Novak, Kunes, Chaput,
  and Pickett}}]{Novak-2006}
\bibinfo{author}{\bibfnamefont{P.}~\bibnamefont{Novak}},
  \bibinfo{author}{\bibfnamefont{J.}~\bibnamefont{Kunes}},
  \bibinfo{author}{\bibfnamefont{L.}~\bibnamefont{Chaput}}, \bibnamefont{and}
  \bibinfo{author}{\bibfnamefont{W.~E.} \bibnamefont{Pickett}},
  \bibinfo{journal}{Physica Status Solidi B: Basic Solid State Physics}
  \textbf{\bibinfo{volume}{243}}, \bibinfo{pages}{563} (\bibinfo{year}{2006}).

\bibitem[{\citenamefont{Aichhorn}()}]{Aichhorn-2013}
\bibinfo{author}{\bibfnamefont{M.}~\bibnamefont{Aichhorn}},
  \bibinfo{note}{private communication}.

\end{thebibliography}

\end{document}